\documentclass[aps,pre,groupedaddress,floatfix,showpacs]{revtex4-1}

\usepackage{amsmath,amssymb}
\usepackage{graphicx}
\usepackage{wrapfig}

\usepackage[T1]{fontenc}
\usepackage[latin9]{inputenc}
\usepackage{amsbsy}
\usepackage{graphicx}
\usepackage[english]{babel}

\begin{document}

\title{Brownian Dynamics Simulation of Polydisperse Hard Spheres}

\author{A. Scala}
\affiliation{ISC-CNR Dipartimento di Fisica, Sapienza Universit\`a di Roma
Piazzale Moro 5, 00185 Roma, Italy}
\affiliation{IMT Alti Studi Lucca, piazza S. Ponziano 6, 55100 Lucca, Italy}
\affiliation{London Institute of Mathematical Sciences, 22 South Audley St
Mayfair London W1K 2NY, UK}

\date{\today}

\begin{abstract}
Standard algorithms for the numerical integration of the Langevin
equation require that interactions are slowly varying during to the
integration time-step. This in not the case for hard-body systems,
where there is no clear-cut between the correlation time of the noise
and the time-scale of the interactions. Starting from a short time
approximation of the Smoluchowsky equation, we introduce an algorithm
for the simulation of the over-damped Brownian dynamics of polydisperse hard-spheres in absence of hydrodynamics interactions
and briefly discuss the extension to the case of external drifts.
\end{abstract}

\maketitle

\section{Introduction}

The discovery that suspensions of colloidal particles can be tuned
to be excellent experimental realizations of the idealised hard-sphere
(HS) system \cite{Pusey1986,Yethiraj2003} has triggered in the last
decade a renewed interest in the theory and simulation of hard-spheres.
Since colloidal hard spheres have a radius in the size range of $10\sim1000\, nm$,
they can be at the same time much bigger than the solute particles
and small enough to have enough thermal energy to disregard
gravitation; hence colloidal HS dynamics can be modelled as Brownian
motion in presence of hydrodynamic interactions. From the theoretical
point of view, hydrodynamics force are often disregarded and the simple
model of Brownian HSs is employed to understand real HS suspensions;
yet, even for the simple Brownian model only approximate theories
are possible and simulations are needed to discriminate among them.

Brownian dynamics (BD) algorithms integrate numerically Langevin equations;
a common requirement of such algorithms is that interactions in the
system should vary little during an integration time-step. Under such
assumption, particle displacements are calculated
keeping forces constant during the integration time-step \cite{ATBook}.
In the case of continuous potentials, computational efficiency worsens
as the interactions become steeper. In the extreme case of hard-body
interactions, stochastic calculus is not naively applicable and standard
numerical integrators become ill defined. On the other hand, the Kramer's
equation \cite{Kramers1940} associated with Brownian motion is well
defined when suitable boundary conditions taking account of stepwise
interaction are implemented \cite{EDlang4HS}. One general strategy
to develop numerical integrators for stochastic differential equations
is to work on the short time expansion (in particular on Trotter expansions
\cite{Trotter59}) of the associated Fokker-Plank equation \cite{DeFabritiis2006}.
We will develop an approach similar in spirit \cite{BD4HS}, by coming
to integrate the over-damped Brownian dynamics (OBD) with stepwise
interactions via suitable approximations of the associated Smoluchowsky
equation (SE).

In section \ref{sec:OBD} we recapitulate and justify the standard event 
driven BD algorithms for homogeneous systems of HSs; in section \ref{sec:POLY} 
we extend such schemes to the case of polydispersity and in section \ref{sec:DRIFT} 
we analyse the extension to the case of constant drifts.

\section{Overdamped Brownian dynamics\label{sec:OBD}}

The stochastic differential equation describing an homogeneous system
of overdamped Brownian particles is 
\[
\partial_{t}\vec{r}_i=\vec{f}_i+\vec{\xi}_i
\]
where $\vec{r}_{i}$ are the coordinates of the $i^{th}$ particle,
$\vec{f}_{i}$ are the non-random forces acting on $i$ and $\vec{\xi}_{i}$
is an uncorrelated Gaussian noise
\[
\left(\vec{\xi}_i\otimes\vec{\xi}_j\right)_{\alpha\beta}=2D\delta_{\alpha\beta}
\]
whose amplitude is twice its diffusion coefficient $D$. In the case
of HSs of diameter $\sigma$, the force $f_{i}$ contains infinite
impulsive contributions due to the HS interaction potential
\[
V_{ij}=\left\{ \begin{array}{c}
0\,\,\,\, for\,\left|\vec{r}_{ij}\right|>\sigma\\
\infty\,\, for\,\left|\vec{r}_{ij}\right|\leq\sigma
\end{array}\right.
\]

Such a force is not Lipschitz continuous and standard methods for
stochastic differential equations become out of reach \cite{KloedenPlatenBook,KannanBook}.
Notice that already for systems of classical particles of mass $m$ in the 
micro-canonical ensemble, the HS force take a peculiar velocity-dependent form
\[
\vec{f}_{ij}dt=-m\vec{v}_{ij}\delta\left(\left|\vec{r}_{ij}\right|-\sigma\right)
\]
where $\vec{v}_{ij}$ is the relative velocity between particles
$i$ and $j$ along the direction of $\vec{r}_{ij}=\vec{r}_{i}-\vec{r}_{j}$; 
in the case of overdamped Brownian HSs, velocities are not defined and the HS
conditions $\left|\vec{r}_{ij}\right|\geq\sigma$ must be interpreted
as boundary conditions. In fact, the Fokker-Plank equation associated
to the OBD of HSs takes the very simple form of a free Smoluchowsky
equation \cite{Smoluchowsky1916} 
\begin{equation}
\partial_{t}P\left( \mathbf{r},t\right)=D\nabla^{2}P\left(\mathbf{r} ,t\right)\label{eq:Smoluchowsky}
\end{equation}
with suitable boundary conditions; here $P\left(\mathbf{r},t\right)$
is the probability distribution function (PDF) for the positions 
$\mathbf{r}=\left\{ \vec{r}_{i}\right\}$.
It is an equation of the form of a divergence
$\partial_{t}P=div\left(\mathbf{j}\right)$
in the current $\mathbf{j}=D\partial_{\mathbf{r}}P$
with $div\left(\mathbf{j}\right)=\partial_{\mathbf{r}}\cdot\mathbf{j}$;
all the complexity is in the implementation of the hard-sphere
impenetrability by a reflecting (zero current) condition 
\[
\left.\hat{\mathbf{n}}\cdot\mathbf{j}\right|_{\partial\Omega}=0
\]
on the time-dependant boundary $\partial\Omega$ corresponding to
$\left|\vec{r}_{ij}\left(t\right)\right|=\sigma$ (i.e. spheres
$i$ and $j$ are at contact at time $t$); here 
and $\hat{\mathbf{n}}$ is the normal to $\partial\Omega$.

To build up an algorithm to integrate such a system, one has to rely
on physical intuition: considering integration steps $\Delta t$
small enough, particles will perform on average free random walks until 
some couples of particles are  \textquotedbl{}near enough\textquotedbl{} to interact. 
This is the basis of many algorithms
for OBD in the case of HSs: first, independent particles displacements
are extracted according to the free Green's function for single particle
diffusion
\begin{equation}
G^{free}_1\left(\vec{r},t+\Delta t|\vec{r}_0,t\right)
\propto\exp\left[-\left(\vec{r}-\vec{r}_0\right)^{2}/2D\Delta t\right]\label{eq:free1pGreen}
\end{equation}; 
then, overlaps are taken account to correct such displacements \cite{Cichocki90sim,Sillescu94,Strating99,Barenbrug02,Terada2001,Foffi05,BD4HS}.
In all such schemes, the implicit assumption is that for small time-steps $\Delta t$
the evolution of the full $P\left( \mathbf{r},t\right)$ factorizes either in single 
particle free evolutions $p\left( \vec{r}_i,t\right)$ or in the evolution 
$p\left( \vec{r}_i,\vec{r}_i,t\right)$ of two interacting particles. 

As shown in \cite{Strating99}, naively chosen corrections can lead
to the wrong dynamics. For purely HS interactions, a naive algorithm
\cite{Terada2001,Foffi05} that transforms the displacements $\Delta\vec{r}_{i}$
in fictive velocities $\boldsymbol{v}_{i}=\Delta\vec{r}_{i}/\Delta t$
and evolves the system according to the rule of standard event-driven
molecular dynamics \cite{RapaBook} (EDMD) has been shown to approximate
correctly the SE of the system \cite{BD4HS}.
In such an approach, the time step $\Delta t$ is fixed; at each time-step,
the velocities of the particles are extracted according to the Maxwell
distribution at a fictive temperature $T$ and a fully fledged EDMD
simulation \cite{RapaBook} is performed between time $t$ and $t+\Delta t$.
The temperature $T$ is chosen such that the average displacement
in absence of collisions is exactly eq.\ref{eq:free1pGreen}.

\begin{figure}
\begin{centering}
\includegraphics[width=0.8\columnwidth]{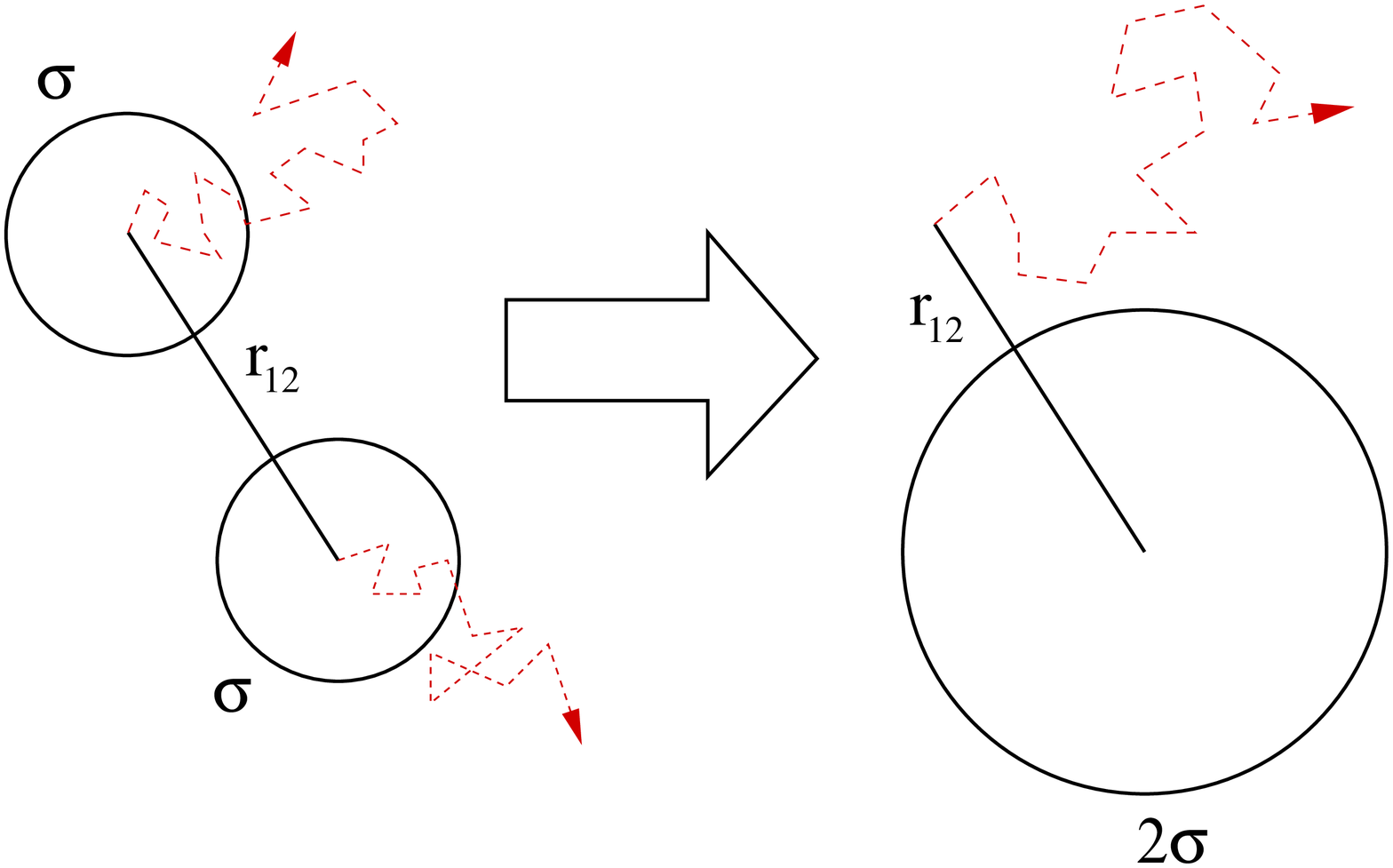}
\par\end{centering}
\caption{The interaction among two Brownian HSs of diameter $\sigma$ can be mapped to the 
solvable problem of a Brownian point particle moving in the presence of an
reflecting boundary given by a sphere of radius $\sigma$.}
\label{fig:twoHSs2PPwithSphBound}
\end{figure}

In order to justify such algorithms, several hypothesis must be
done. First, the time step $\Delta t$ must be small enough that only
binary collisions must be relevant, i.e. the average displacement 
$\left\langle \left|\Delta\vec{r}_{i}\right|\right\rangle \sim\Delta t^{1/2}$
must be much smaller than the average inter-particle distance 
\begin{equation}
\left\langle \left|\Delta\vec{r}_{i}\right|\right\rangle \ll\rho^{-1/d}-\sigma\label{eq:BinCollCond}
\end{equation}; 
here $\rho$ is the number density and $d$ is the dimension of the system. 
In such a limit, the interaction
among two overdamped Brownian HSs $i$ and $j$ can be mapped to the
problem of a point overdamped Brownian particle in presence of a sphere
by the change of coordinates (fig. \ref{fig:twoHSs2PPwithSphBound})
\begin{equation}
\left\{ \begin{array}{ccc}
\vec{r}_{ij} & = & \vec{r}_{i}-\vec{r}_{j}\\
\vec{R}_{CM} & = & \left(\vec{r}_{i}+\vec{r}_{j}\right)/2
\end{array}\right.\label{eq:homCMtransform}
\end{equation}; 
in such a reference system, the Brownian center of mass $\vec{R}_{CM}$ is subject 
to free diffusion $\partial_{t}\vec{R}_{CM}=\Xi$ while $\vec{r}_{ij}$ satisfies 
the SE with spherical boundary conditions
\begin{equation}
\left\{ \begin{array}{ccc}
\partial_{t}\vec{r}_{ij} & = & \vec{\xi}\\
\left|\vec{r}\right| & \geq & \sigma
\end{array}\right.\label{eq:OBD1sphere}
\end{equation};
here $\vec{\xi}=\vec{\xi}_i-\vec{\xi}_j$
and $\vec{\Xi}=\left(\vec{\xi}_i+\vec{\xi}_j\right)/2$.
Notice that $\vec{\xi}_{ij}$ and $\Xi$ are mutually orthogonal 
Gaussian noises and therefore the equations for $\vec{R}_{CM}$
and for $\vec{r}_{ij}$ can be solved independently. 
Equation (\ref{eq:OBD1sphere}) can be exactly solved \cite{Hanna82,Ackerson82} 
but the solution is in the form of 
an infinite sum in the Laplace domain; therefore, further approximations are 
needed as it is not suitable for the fast for numerical
implementations necessary to simulate many-body system. 
In particular, the condition of small displacements during the time step $\Delta t$
can be pushed to satisfy also an additional ``\emph{flat wall}'' condition
\begin{equation}
\left\langle \left|\Delta\vec{r}_{i}\right|\right\rangle \ll\sigma\label{eq:FlatWallCond}
\end{equation}

\begin{figure}
\begin{centering}
\includegraphics[width=0.6\columnwidth]{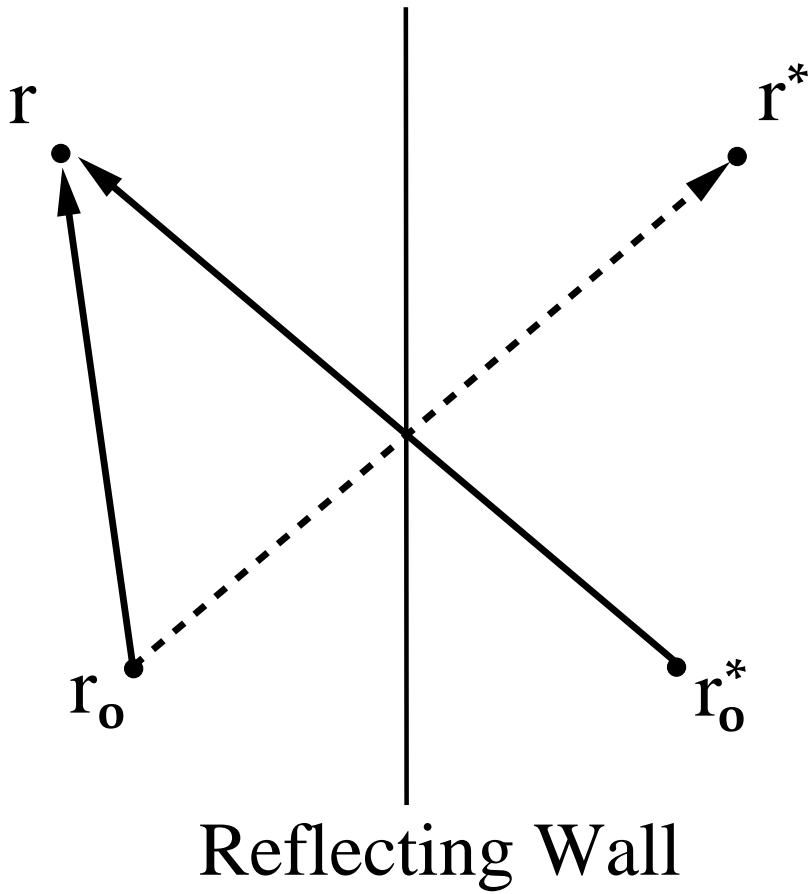}
\par\end{centering}
\caption{Image method of solution for the Smoluchowsky equation in presence of a flat reflecting boundary.
The probability of an overdamped Brownian particle of reaching a 
point $\vec{r}$ at time $t$ is given 
by the sum of the probability $G_1^{free}\left(\vec{r},t+\Delta t | \vec{r}_0 , t\right)$ that the particle 
goes from $\vec{r}_0$ to $\vec{r}$ by free diffusion plus the probability 
$G_1^{free}\left(\vec{r},t+\Delta t | \vec{r}^*_0 , t\right)$ that the particle goes from 
the image $\vec{r}^*_0$ of the initial point to the same $\vec{r}$ also by free diffusion. 
The latter is equal to the probability 
$G_1^{free}\left(\vec{r}^*,t+\Delta t | \vec{r}_0 , t\right)$ that the particle goes by free 
diffusion from  $\vec{r}_0$ to the image $\vec{r}^*$ of the point $\vec{r}$. Therefore, to 
implement numerically the image method, it suffices to 
implement the following algorithm:
$(1)$ extract the final position 
$\vec{r}$ according to the solution $G_1^{free}$ of the free Smoluchowsky equation and 
$(2)$ reflect $\vec{r}$ if it goes beyond the 
hard boundary.
}
\label{fig:Image-method}
\end{figure}

In such a situation, binary collisions modelled by eq.(\ref{eq:OBD1sphere}) 
happen on average only between particles at an initial distance
$\vert\vec{r}_ij\vert\cong\sigma$, i.e. at distances much smaller than the radius of 
curvature of the spherical boundary. In such a situation the boundary can be 
approximated as a  flat wall. Shifting to Cartesian coordinates such that 
the origin lies on the intersection of $\vec{r}_{ij}$ with boundary 
and orienting the $y$,$z$ axis tangentially to the surface, 
the system 
factorizes in two free Smoluchowsky equations for the $y$,$z$ coordinates and 
a one dimensional equations for the $x=x_{ij}-\sigma$ coordinate 
\[
\left\{ \begin{array}{ccc}
\partial_{t}x & = & \left(\vec{\xi}\right)_x\\
x & \leq & 0
\end{array}\right.
\]
which can be exactly solved \cite{Smoluchowsky1916} 
with the image method \cite{RednerBOOK2001}
(fig.\ref{fig:Image-method}):
\[
G_{1}^{wall}\left(\vec{r}\right)\propto\left\{ \begin{array}{ccc}
e^{\frac{-|\vec{r}-\vec{r}_{0}|^{2}}{2D\Delta t}}+e^{\frac{-|\vec{r}+\vec{r}_{0}|^{2}}{2D\Delta t}} & for & x\leq0\\
0 & for & x>0
\end{array}\right.
\].

The whole solution $G_{1}^{wall}$ for a point particle
starting in $\vec{r}_{0}$ consists of the superposition
in the $x\leq0$
semi-space of the free Green's function (\ref{eq:free1pGreen}) of
a particle in $\vec{r}_{0}=\left(x_{0},y_{0},z_{0}\right)$ and
an image particle in $\vec{r}_{0}^{*}=\left(-x_{0},y_{0},z_{0}\right)$. 
Such a solution can be implemented with just a single operation
by extracting the new position $\vec{r}\left(t+\Delta t\right)$
according to (\ref{eq:free1pGreen}) and reflecting the $x$ coordinate
whenever $x>0$. An event-driven algorithm implementing such scheme is the following:
\begin{enumerate}
\item extract the random displacements $\Delta x$, $\Delta y$, $\Delta z$,
\item define a fictive velocity $v_{x}=\Delta x/\Delta t$
\item calculate the fictive collision time $t_c:x_{0}+v_{x}t_c=0$
\item calculate fictive the post-collision velocity $v_{x}^{*}=-v_{x}$
\item calculate the final position $x\left(t+\Delta t\right)=x\left(t\right)+v_{x}\cdot t_c+v_{x}^{*}\cdot\left(\Delta t-t_c\right)$
\end{enumerate}

When mapping back from the Brownian Center of Mass (BCoM) reference system to 
the original particles' coordinates, the whole Brownian collision 
$\vec{r}_{i}\left(t\right)\rightarrow\vec{r}_{i}\left(t+\Delta t\right)$,
$\vec{r}_{j}\left(t\right)\rightarrow\vec{r}_{j}\left(t+\Delta t\right)$
follows a procedure strictly recollecting the collision of two classical HSs:
\begin{enumerate}
\item extract two random displacements $\Delta\vec{r}_{i}$, $\Delta\vec{r}_{j}$
according to (\ref{eq:free1pGreen})
\item define two fictive velocities $\vec{v}_{i}=\Delta\vec{r}_{i}/\Delta t$,
$\vec{v}_{j}=\Delta\vec{r}_{j}/\Delta t$ 
\item calculate the fictive collision time $t_c\in\left[0,\Delta t\right]$,
and the normal $\hat{n}^*$ between the two spheres at contact
at time $t_c$ 
\item calculate the fictive post-collision velocities $\vec{v}_{i}^{*}=\vec{v}_{i}-2\left(\hat{n}^{*}\cdot\vec{v}_{ij}\right)\hat{n}^{*}$
and
$\vec{v}_{j}^{*}=\vec{v}_{j}+2\left(\hat{n}^{*}\cdot\vec{v}_{ij}\right)\hat{n}^{*}$
with $\vec{v}_{ij}=\vec{v}_{i}-\vec{v}_{j}$
\item calculate the final positions $\vec{r}_{i}\left(t+\Delta t\right)=\vec{r}_{i}\left(t\right)+\vec{v}_{i}\cdot t_c+\vec{v}_{i}^{*}\cdot\left(\Delta t-t_c\right)$
and
$\vec{r}_{j}\left(t+\Delta t\right)=\vec{r}_{j}\left(t\right)+\vec{v}_{j}\cdot t_c+\vec{v}_{j}^{*}\cdot\left(\Delta t-t_c\right)$
\end{enumerate}

Therefore, any event-driven molecular dynamic code \cite{RapaBook}
can be adapted to simulate the OBD of HSs by extracting at each step
$\Delta t$ the velocities of the particles according to independent
Gaussian distributions such that $\left\langle \vec{v}_{i}^{2}\right\rangle =2dD\Delta t^{-1}$;
a consistency check  to perform is to ensure that within 
the chosen time-step $\Delta t$ less than one collision 
per particle occurs on average.

\section{Polydispersity and Brownian Collisions\label{sec:POLY}}

In realistic colloidal system there is an inherent polydispersity
in particle size; more generally, one could be also interested to
mixtures of HSs with different characteristics as in all the 
studies where crystallization must be avoided. 
Let us now suppose
that our system is composed of HSs of diameters $\sigma_{i}$ subject
to overdamped Brownian motion with particle-dependent
free diffusion coefficient $D_{i}$, i.e. to noises of amplitude 
$\langle\vec{\xi}_{i}^{2}\rangle=2dD_{i}$; 
as usual, trajectories
are subject to no-flux boundary constraints 
$\left|\vec{r}_{i}(t)-\vec{r}_{j}(t)\right|\geq \sigma_{ij}$
where $\sigma_{ij}=\left(\sigma_{i}+\sigma_{j}\right)/2$.
Let's suppose again to fix a time-step $\Delta t$ small enough to
consider only binary collisions; 
for two particles $i$ and $j$ the equations become 
\begin{equation}
\begin{array}{c}
\partial_t \vec{r}_i = \vec{\xi}_i \\
\partial_t \vec{r}_j = \vec{\xi}_j \\
\vert \vec{r}_{ij} \vert \geq \sigma_ij
\end{array}
\label{eq:2polySDE}
\end{equation}.

The first step is to separate equations (\ref{eq:2polySDE});
transformation (\ref{eq:homCMtransform}) does not succeed 
any longer as the transformed noises have a non-zero correlation 
$D_{i}-D_{j}$. To properly define the ``Brownian center of mass''
(BCoM), we start from the ansatz 
$\vec{R}_{CM}=a_{i}\vec{r}_{i}+a_{j}\vec{r}$ 
and impose zero correlation among the random displacements of 
the BCoM and the inter-particle distance: 
$0=\left\langle \Delta \vec{R}_{CM}\Delta \vec{r}_{ij}\right\rangle 
\propto a_{i}D_{i}-a_{j}D_{j}$;
a proper dimensionless choice is 
$a_{i}=(D_{i}+D_{j})/D_{j}\propto D_{i}^{-1}$.

In the limit of $\Delta t$ small enough such that the 
$\vert \vec{r}_{ij}\vert=\sigma_{ij}$ the boundary 
can be approximated  with a flat hard wall and the zero flux 
problem can be solved along the dimension perpendicular to the 
wall (we assume it is the $x$ direction). Let's again define 
fictive velocities both in the original reference system 
and in the BCoM system $v_{\alpha}=\Delta x_{\alpha}/\Delta t$
for $\alpha\in\left\{ i,j,CM,ij\right\} $.

The change of coordinates to the BCoM system is

\[
\left(\begin{array}{c}
v_{CM}\\
v_{ij}
\end{array}\right)=A\left(\begin{array}{c}
v_{i}\\
v_{j}
\end{array}\right)=\left[\begin{array}{cc}
\frac{D_{i}+D_{j}}{D_{i}} & \frac{D_{i}+D_{j}}{D_{j}}\\
1 & -1
\end{array}\right]\left(\begin{array}{c}
v_{i}\\
v_{j}
\end{array}\right)
\]

with inverse

\[
\left(\begin{array}{c}
v_{i}\\
v_{j}
\end{array}\right)=A^{-1}\left(\begin{array}{c}
v_{CM}\\
v_{ij}
\end{array}\right)=\left[\begin{array}{cc}
\frac{D_{i}D_{j}}{\left(D_{i}+D_{j}\right)^{2}} & -\frac{D_{i}}{D_{i}+D_{j}}\\
\frac{D_{i}D_{j}}{\left(D_{i}+D_{j}\right)^{2}} & \frac{D_{j}}{D_{i}+D_{j}}
\end{array}\right]\left(\begin{array}{c}
v_{CM}\\
v_{ij}
\end{array}\right)
\]

In the BOcM system the collision corresponds to imposing the 
no-flux boundary collision and is simply

\[
\left(\begin{array}{c}
v'_{CM}\\
v'_{ij}
\end{array}\right)=\left[\begin{array}{cc}
1 & 0\\
0 & -1
\end{array}\right]\left(\begin{array}{c}
v_{CM}\\
v_{ij}
\end{array}\right)
\]

and therefore the fictive velocities after the collision are

\[
\left(\begin{array}{c}
v_{i}^{*}\\
v_{j}^{*}
\end{array}\right)=A^{-1}\left(\begin{array}{c}
v_{CM}^{*}\\
v_{ij}^{*}
\end{array}\right)=A^{-1}\left[\begin{array}{cc}
1 & 0\\
0 & -1
\end{array}\right]\left(\begin{array}{c}
v_{CM}\\
v_{ij}
\end{array}\right)=A^{-1}\left[\begin{array}{cc}
1 & 0\\
0 & -1
\end{array}\right]A\left(\begin{array}{c}
v_{i}\\
v_{j}
\end{array}\right)
\]

with collision matrix

\[
C_{Brown}=A^{-1}\left[\begin{array}{cc}
1 & 0\\
0 & -1
\end{array}\right]A=\left[\begin{array}{cc}
\frac{D_{j}-D_{i}}{D_{i}+D_{j}} & \frac{2D_{i}}{D_{i}+D_{j}}\\
\frac{2D_{j}}{D_{i}+D_{j}} & \frac{D_{i}-D_{j}}{D_{i}+D_{j}}
\end{array}\right]
\]

This is to be compared with the classical collision matrix for two
elastic particles of masses $m_{i}$, $m_{j}$

\[
C_{class}=\left[\begin{array}{cc}
\frac{m_{i}-m_{j}}{m_{i}+m_{j}} & \frac{2m_{j}}{m_{i}+m_{j}}\\
\frac{2m_{i}}{m_{i}+m_{j}} & \frac{m_{j}-m_{i}}{m_{i}+m_{j}}
\end{array}\right]
\]

that has a similar structure if one but with switched indexes 
$i\leftrightarrow j$ such the role of the mass $m_{i}$ during 
a collision is played by the inverse diffusivity $D_{i}^{-1}$.

As a check, we consider a fixed particle with $D_{i}=0$; this is
equivalent for particle $i$ as having an infinite mass (noise does
not move it) and the collision is 

\[
\left(\begin{array}{c}
v_{i}^{*}\\
v_{j}^{*}
\end{array}\right)=C\left(\begin{array}{c}
v_{i}\\
v_{j}
\end{array}\right)=\left[\begin{array}{cc}
1 & 0\\
2 & -1
\end{array}\right]\left(\begin{array}{c}
v_{i}\\
v_{j}
\end{array}\right)=\left(\begin{array}{c}
v_{i}\\
2v_{i}-v_{j}
\end{array}\right)
\]

as it should be (notice that $v_{j}^{*}=-v_{j}$ as $v_{i}=0$ for
$D_{i}=0$); an analogous result comes by sending the ``mass'' of
particle $j$ to zero (i.e. $D_{j}=\infty$). 

To summarize, let's recall that the full event-driven collision scheme for classical particles is
\begin{enumerate}
\item calculate collision time $t_c$ from the ``good'' root of 
$\left\Vert \vec{r}_{ij}+\vec{v}_{ij}t_c\right\Vert =\vec{\sigma}_{ij}$
\item bring particles at contact $\vec{r}_{i}=\vec{r}_{i}+\vec{v}_{i}t_c$
,$\vec{r}_{j}=\vec{r}_{j}+\vec{v}_{j}t_c$ 
\item let $\vec{\sigma}_{ij}=\vec{r}_{ij}(t_c)$ , $\hat{\sigma}_{ij}=\vec{\sigma}_{ij}/\left\Vert \vec{\sigma}_{ij}\right\Vert $
,$\mathit{v}_{i}=\vec{v}_{i}\cdot\hat{\sigma}_{ij}$,$\mathit{v}_{j}=\vec{v}_{j}\cdot\hat{\sigma}_{ij}$
\item pre-collision: $\left(\begin{array}{c}
\mathit{v}'_{i}\\
\mathit{v}'_{j}
\end{array}\right)=C_{class}\left(\begin{array}{c}
\mathit{v}_{i}\\
\mathit{v}_{j}
\end{array}\right)$
\item collision: $\vec{v}_{i}'=\vec{v}_{i}-\mathit{v}{}_{i}\hat{\sigma}_{ij}+\mathit{v}'_{i}\hat{\sigma}_{ij}$
, $\vec{v}_{j}'=\vec{v}_{j}-\mathit{v}{}_{j}\hat{\sigma}_{ij}+\mathit{v}'_{j}\hat{\sigma}_{ij}$ 
\end{enumerate}
Therefore, to modify an Event Driven code for polydisperse HSs
into an Event Driven Brownian Dynamics, 
simply extract random displacements $\Delta\vec{r}_{i}$ 
at fixed intervals $t,t+\Delta t,t+2\Delta t,\ldots$, 
define fictive velocities $\vec{v}_{i}=\Delta\vec{r}_{i}/\Delta t$
and evolve the system for a time $\Delta t$ using for the collision
the matrix $C_{Brown}$ instead of $C_{class}$.

\section{Constant drifts\label{sec:DRIFT}}

Insofar, only system not subject to external forces have 
been considered.
For overdamped Brownian motion, constant forces add 
constant drifts to the random displacements
of the particles. 
Polydisperse particles are to be expected to experience drifts
of different magnitudes even in presence of an homogeneous fields (like gravity or an electrical field). 
Therefore, the two body equation becomes 

\begin{equation}
\left\{ \begin{array}{ccc}
\partial_{t}\vec{r}_{ij} & = & \vec{\xi}_{ij}+\vec{g}_{ij}\\
\left|\vec{r}_{ij}\right| & \geq & \left(\sigma_{i}+\sigma_{j}\right)/2
\end{array}\right.\label{eq:OBD1drift}
\end{equation}
where $\vec{g}_{ij}=\vec{g}_{i}-\vec{g}_{j}$ is the difference
among the constant drifts of the two particles. 
Notice that homogeneous drifts do not produce
any change in the equations for the inter-particles distances $\vec{r}_{ij}$ as $\vec{g}_{ij}=0$,
but just add a constant drift to the BCoM $\vec{R}_{CM}$; 
therefore, Brownian Event Driven simulations can be implemented 
extracting particle displacements according to the  
Green function of the Smoluchowsky equation for a 
single particle with drift instead of (\ref{eq:free1pGreen}).

To see the effects of no-zero $\vec{g}_{ij}$, 
let's consider again a $\Delta t$ small enough 
such that the collision can be approximatively by a
flat wall. Factorizing the motion in the directions perpendicular 
($x$) and parallel ($y,z$) to the wall, one is left with solving 
the Smoluchowsky equation with a reflecting boundary in the case 
of constant drift.
This problem has been solved at the beginning of the last century
in the seminal paper by M. Smoluchowsky \cite{Smoluchowsky1916} (pp.
569-574; see \cite{LammJCP1983}, pp. 2714 for an English version).
Assuming that the reflecting boundary is the plane
$\partial\Omega=\left\{ x=0\right\} $, 
the evolution of the probability distribution function 
follows the equation

\[
\left\{ \begin{array}{cc}
 & \partial_{t}p=D\partial_{x}^{2}p+c\partial_{x}p\\
subject\,\, to & \left.\left(\partial_{x}+c\right)p\left(x,t\right)\right|_{x=0}
\end{array}\right.
\]

where $c=-\beta g$, $\beta=1/k_{B}T$ is the inverse temperature
and $g$ is a constant force.

Solving for the initial condition 
$p\left(x,t=0|x_{0}\right)=\delta\left(x-x_{0}\right)$
one obtains the solution

\begin{equation}
\begin{array}{c}
p\left(x,t|x_{0}\right)=\frac{1}{2\sqrt{\pi Dt}}\left[e^{-\frac{\left(x-x_{0}\right)^{2}}{4Dt}}+e^{-\frac{\left(x+x_{0}\right)^{2}}{4Dt}}\right]e^{-\frac{c\left(x-x0\right)}{2D}-\frac{c^{2}t}{4D}}\\
+\frac{c}{D\sqrt{\pi}}e^{-\frac{cx}{D}}erfc\left(\frac{x-x_{0}-ct}{\sqrt{4Dt}}\right)
\end{array}+
\label{eq:SmolDrift1d}
\end{equation}
where $erfc\left(z\right)=1-erf\left(z\right)$ is the complementary
error function and the
error function is $erf(z)=\sqrt{4/\pi}\int_{-\infty}^{z}ds\exp\left[-s^{2}\right]$.

\begin{figure}
\begin{centering}
\includegraphics[width=0.8\columnwidth]{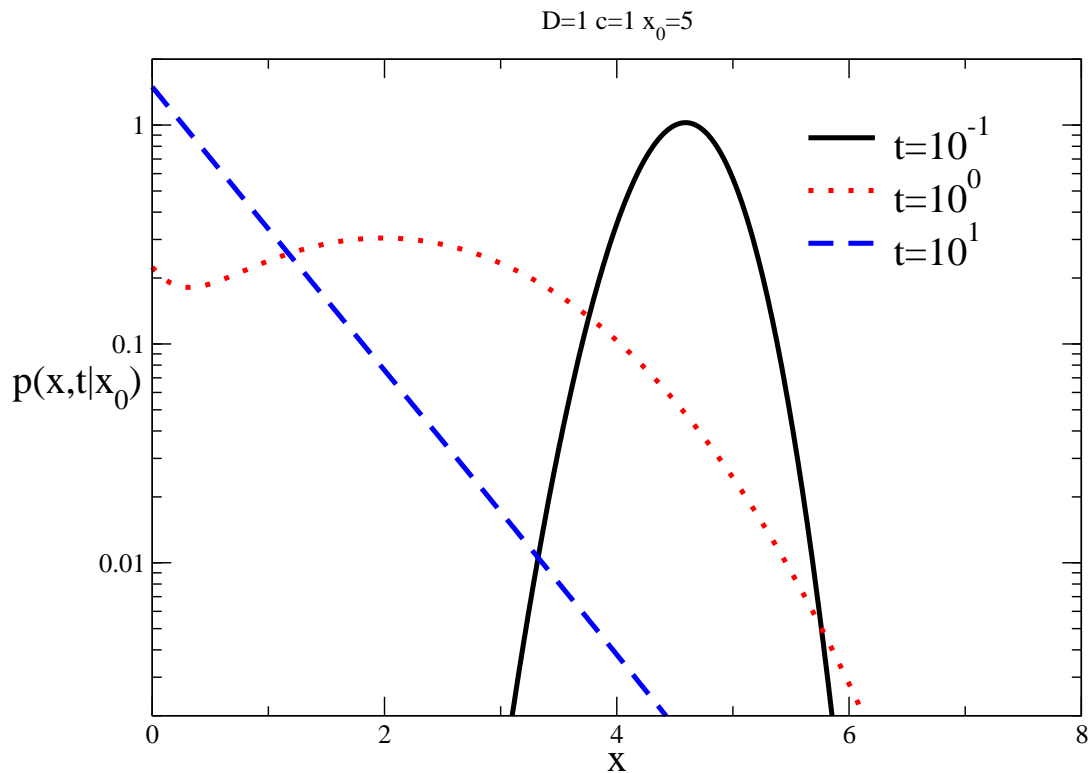}
\par\end{centering}

\caption{Plot of particular solutions of the Smoluchowsky 
equation with drift in presence of a reflecting boundary 
for diffusion $D=1$, drift $c=1$ and initial position $x_{0}=1$;
all parameters   are expressed in dimensionless form, 
in which physical properties are scaled using appropriate 
combinations of the characteristic size, energy and/or mass . 
Notice that, while at short time the probability distribution 
is  essentially a shifting Gaussian  (a parabola in the 
log-linear plot), the probability accumulates at latter times 
near the boundary while approaching at long times the Boltzmann equilibrium corresponding to an exponential solution.
}
\label{fig:SmolG}
\end{figure}

Such a solution is not amenable of a simple geometric implementation
in terms of a naive collision mechanism: in this case, 
for each collision
one must transform the coordinates to the BCoM reference system, 
extract the displacements according to (\ref{eq:SmolDrift1d}) 
and transform back to the
original coordinate system. As a further caveat, high enough 
constant inter-particle drifts imply the accumulation 
of particles at short distances (fig.\ref{fig:SmolG}); 
the appearance of such inhomogeneous structures is a critical
situation is critical for event-driven algorithms 
as it can produce unacceptable slowing-down of the simulation
(via the growth of the number of collision per unit time) 
and eventually numerical errors \cite{RapaBook}. 
Notice that such issues of non-zero drifts among nearby HSs 
are often disregarded in the simulations 
of sheared particles where collisions between hard disks 
or hard spheres are implemented as elastic collisions of 
the fictive velocities
\cite{Strating99,Henrich2009,KrugerPRE2010,KrugerEPJE2011,MarechalJCP2011}
. 
Since for HSs structural quantities like the pressure 
are strictly related to the radial distribution function 
at contact, a careful analysis of the importance of the 
drift term in relation to the strength of the noise should be
performed to avoid disregarding possible relevant corrections.

\acknowledgments

The author thanks Th. Voigtmann his hospitality at the physics department of Konstanz where this work has been conceived during 
the long, useful discussions with him and his students.
The author acknowledges the support of the 
CNR-PNR National Project Crisis-Lab.

\bibliographystyle{plain}
\bibliography{LANG4HS}

\end{document}